# Effects of Mn-substitution for Co on the structure and physical properties of $Na_\gamma CoO_2$

W. Y. Zhang,[1] H. C. Yu,[2] Y. G. Zhao,[1]* X. P. Zhang,[1] Y. G. Shi,[2] Z. H. Cheng,[2] and J. Q. Li[2]

[1] *Department of Physics, Tsinghua University, Beijing 100084, P. R. China*

[2] *Institute of Physics, Chinese Academy of Sciences, Beijing 100080, P. R. China*

**Abstract**

Structure, transport and magnetic properties have been investigated for layered $NaCo_{1-x}Mn_xO_2$ ($0 \leq x \leq 0.5$) materials. With the increase of Mn-doping amount, measurements of x-ray diffraction indicate that the in-plane lattice parameter *a* decreases progressively, and, conversely, the lattice parameter *c* increases rapidly. Small amount of Mn doping (i.e. 3% Mn in our work) could result in a metal-insulator transition. Electron energy loss spectroscopy (EELS) measurement shows that the Co valence is about 3.3 and remains unchanged with doping. In contrast, the valence of Mn increases rapidly with doping and reaches $3.7 \pm 0.1$ for the sample with x=0.5. The temperature dependence of magnetization of the samples is found to obey the Curie-Weiss law. Some relevant results have been discussed in terms of the disorder effect induced by doping of magnetic ions.

PACS numbers: 72.80.Ga; 71.27.+a; 74.62.Dh

**Introduction**

Layered $Na_\gamma CoO_2$ compound has attracted considerable attention due to its unique structure and physical properties, e.g., large room temperature Seebeck coefficient 100 μv/K which is nearly ten times higher than that of typical metals [1-3], and existence of superconductivity of water-intercalated $Na_xCoO_2$ compound which is a breakthrough in the search for new layered transition metal oxide superconductors [4,5]. Layered $Na_xCoO_2$ compound has a two-dimensional (2D) crystal structure where Na ion and $CoO_2$ block are alternately stacked along the c axis. Each Co ion is surrounded by slightly distorted oxygen octahedra and the Co ions form the 2D triangular lattice, which may be the realization of Anderson's original triangular lattice RVB system [6]. Charge carrier transport is thought to be restricted mainly to these $CoO_2$ planes, as in the case of the $CuO_2$ planes for the high-$T_c$ cuprates, thus the physical properties are expected to be highly 2D. For example, the resistivity of $Na_xCoO_2$ single crystals shows anisotropy both in magnitude and temperature dependence [1]. However, the 2D triangular lattice of Co ions in $Na_xCoO_2$ is different from the 2D square lattice of cuprates and the effective hopping of 3d electrons between Co ions is reduced since the Co-O-Co bond angle is about $98.5^o$ instead of $180^o$ in cuprates [7, 8].

Evidences are accumulating to show that $Na_\gamma CoO_2$ is a strong correlated system [9-12]. In this kind of systems, spin-charge-orbital interactions are in a subtle balance and dramatic change of physical properties could be induced with a slight variation in, e.g., the carrier density or lattice distortion which is closely related to the composition of the compounds. A number of experimental results show that Na-site substitution has strong influences on the physical properties of $Na_xCoO_2$ compounds [13-15]. Up to now, to our knowledge, little work has been done on the effect of Co-site substitution with magnetic ions on the structure and physical properties of $Na_xCoO_2$ compounds. Terasaki et al studied $NaCo_{2-x}Cu_xO_4$ focusing on their thermopower [16]. It was found that the solid solubility limit of Cu in $NaCo_{2-x}Cu_xO_4$ is very small (x=0.2) and the samples remain metallic.

In this paper, the effects of the substitution of Mn for Co on structure, transport and magnetic properties of layered $Na_xCoO_2$ compounds have been investigated. It is found that the solid solubility of Mn in $Na_xCoO_2$ is very large, at least larger than the maximum doping (x=0.5) in the present work. EELS measures demonstrate that the valence states of manganese change evidently with the increase of the doping level. Small amount of Mn-substitution could induce a metal-insulator transition in $Na_xCoO_2$.

**Experimental**

Polycrystalline samples of $NaCo_{1-x}Mn_xO_2$ (x = 0, 0.03, 0.05, 0.07, 0.1, 0.2, 0.3, 0.4, 0.5) were prepared by solid-state reaction. A stoichiometric amount of $Na_2CO_3$, $Co_3O_4$, or $MnO_2$ was mixed and sintered at 800~860 $^0$C for 14 h in air. The product was finely ground, pressed into pellets, and sintered at 800-900 $^0$C for 10 h in air. Since Na tends to evaporate during calcinations [2], the undoped sample with the starting composition of $NaCoO_2$ is expected to change to about $Na_{0.70}CoO_2$.

The x-ray diffraction was performed using a Rigaku D/max-RB x-ray diffractometer with Cu $K_\alpha$ radiation as the x-ray source in the $\theta$-$2\theta$ scan mode. The electrical resistivity ρ(T) was measured using the four probe method within the temperature range of 5~300 K. Indium was used for the electrical contact. The dc magnetization was measured using a superconducting quantum interference device (SQUID) magnetometer within the temperature range of 5~300 K in a 500 Oe magnetic field.

The TEM investigations were performed on a Tecnai F20 TEM (200kV) equipped with a post column Gatan imaging filter. The energy resolution in the EELS spectra is 0.7 eV under normal operation conditions. In order to minimize the radiation damage under electron beam the samples were cooled below 200K during our TEM observations.

**Results and discussions**

Figure 1 shows the x-ray diffraction patterns of the $NaCo_{1-x}Mn_xO_2$ samples. The Rietveld analysis reveals a hexagonal layered structure with space group of $P6_3/mmc$ and $a = 2.837$ Å, $c = 10.878$ Å for the x = 0 sample. The values of the lattice constant $a$ and $c$ determined here are in good agreement with the standard values of $a = 2.833$ Å and $c = 10.88$ Å obtained by Fouassier et al.[15], suggesting that the sintered samples are mainly composed of $Na_{0.70}CoO_2$ γ phases. With x increasing, no additional peak appears, which shows that Mn is completely substituted for Co. It seems that the degree of solid solution with Mn can be further improved. Figure 2 shows the lattice constant $a$, $c$ and unit cell volume with respect to the Mn doping x. With the increase of doping amount, the lattice parameter $a$ decreases and the lattice parameter $c$ increases. As a result, the unit cell volume of the samples increases with Mn doping. The variation of the lattice parameters $a$, $c$ and unit cell volume, can be understood by considering the valences of Co and Mn obtained from EELS as shown later.

EELS, a powerful technique for material characterization at a nanometer spatial resolution, has been widely used in chemical micro-analysis [17]. In order to obtain the information of the valence states for both Co and Mn in the $NaCo_{1-x}Mn_xO_2$ materials, we have typically performed EELS analyses on x=0.1 and x=0.5 samples, Figure 3(a) shows the EELS spectra of $NaCo_{0.9}Mn_{0.1}O_2$ and $NaCo_{0.5}Mn_{0.5}O_2$ materials taken from an area of about 100 nm in diameter. In these spectra, the typical peaks, i.e. the collective plasmon peaks as well as core edges for Co, Mn, and O elements, are displayed.

For transition metals with unoccupied 3d states, the transition of an electron from 2p state to 3d levels leads to the formation of white lines. The $L_3$ and $L_2$ lines are the transitions from $2p_{3/2}$ to $3d_{3/2}3d_{5/2}$ and from $2p_{1/2}$ to $3d_{3/2}$, respectively. Their intensities are related to the unoccupied states in the 3d bands. In $NaCo_{1-x}Mn_xO_2\cdot$ materials, we have made a series of measurements in association with quantitative analyses by the method as reported by Wang et al [17].

Our first measurement is performed on Co peaks in the spectra, the ratio of $L_3/L_2$ in general is around 2.4, and no evident change occurs along with the increase of the Mn content. These data could yield the Co valence of about 3.3-3.4 for the samples with x=0.1 and 0.5. This valence state is fundamentally in agreement with that obtained for the parent sample $NaCoO_2$ [18].

In contrast to the data obtained for the Co element, a remarkable change of Mn-peaks with the Mn content has been observed in our experiments. Figure 3(b) show a typical spectra for the Mn $L_2$ and $L_3$ peaks obtained from x=0.1 and 0.5, respectively, in both spectra we schematically illustrate the extraction of the intensities for the white lines of the Mn element. The measurements for x=0.1 sample indicate that the ratio of $L_3/L_2$ is about 2.4, which could yield the average of Mn valence of about 3.1-3.3 in this material. The analyses of the x=0.5 sample give rise to the $L_3/L_2$ ratio at around 2.05, moreover, this data could change slightly from one crystalline grain to another. Our systematical analyses conclude that the Mn valence state ranges from 3.6 to 3.8 for the x=0.5 sample.

In accordance with the EELS measurement, the ratio of $Co^{3+}/Co^{4+}$ generally is not equal to that of $Mn^{3+}/Mn^{4+}$ in this kind of materials, instead the ratio of $Mn^{3+}/Mn^{4+}$ changes much quicker than that of $Co^{3+}/Co^{4+}$. Therefore $Mn^{4+}$ ions not only substitute for $Co^{4+}$, but also substitute for some $Co^{3+}$ ions and keep the ratio of $Co^{3+}/Co^{4+}$ constant. In order to keep the charge balance, either the content of Na decreases or the oxygen content increases. Base on the valences of Mn and Co, the variation of lattice parameters *a*, *c* and unit cell volume of $NaCo_{1-x}Mn_xO_2$ can be understood by considering the ion radius of $Co^{3+}$ (0.685 Å), $Co^{4+}$ (0.67 Å), $Mn^{3+}$(0.72 Å) and $Mn^{4+}$(0.67 Å) [19]. The ion radii of $Mn^{3+}$ is comparably larger than that of $Co^{3+}$, $Co^{4+}$ and $Mn^{4+}$. So Mn doping is expected to increase the unit cell volume and the slope will decrease with Mn doping due to the decrease of the $Mn^{3+}/Mn^{4+}$ ratio, consistent with the unit cell volume variation shown in Fig.2. Since $Mn^{3+}$ is a Jahn-Teller ion, distortion with in-plane contraction and out-of-plane (c axis) elongation may occur, leading to the decrease of in-plane lattice parameter *a*.

The variation of resistivity with temperature for $NaCo_{1-x}Mn_xO_2$ is shown in Fig.4. It can be clearly seen that a low level doping with 3% Mn for Co leads to a metal-insulator transition, in contrast to $NaCo_{1-x}Cu_xO_2$ [16]. Since the lattice parameter *c* increases with Mn doping, the coupling between the $CoO_2$ layers decreases and the 2D nature of the $CoO_2$ is more prominent with Mn doping. Generally, it is known that the physical properties of low dimensional systems are very sensitive to disorder, the metal-insulator transition in $NaCo_{1-x}Mn_xO_2$ therefore may be due to the disorder induced Anderson localization. Because $NaCo_{1-x}Cu_xO_2$ remains metallic up to x=0.1, the solid solubility limit of Cu in $NaCoO_2$ [16], the contribution of magnetic disturbance to the $CoO_2$ layers due to Mn doping may play an important role. The inset of Fig.4 presents the dependence of room-temperature resitivity of $NaCo_{1-x}Mn_xO_2$ on Mn content. The resitivity increases gradually with x and an abrupt increase occurs at around 0.07. It has been shown that the distance between Co is very important to determine the electrical property and a critical distance $R_c$ can be used to judge whether $NaMO_2$ (M=Cr, Mn, Fe, Co, Ni) is a metal or an insulator [20, 21]. The same thought may be also useful for the doped systems. The transport mechanism of the doped $NaCoO_2$ needs further study.

Figure 5 shows the result of dc magnetization for the $NaCo_{1-x}Mn_xO_2$ samples. All the samples exhibit Curie-Weiss behavior, indicating the existence of localized moment. Using the Curie-Weiss law $\chi = \chi_0 + C/(T-\theta)$ to fit the experimental data, where $\theta$ is the paramagnetic Curie-Weiss temperature, $\chi_0$ is a sum of temperature-independent terms, and C = $N\mu_{eff}^2/3k_B$ ($\mu_{eff}$ is the effective moment of magnetic ions, $k_B$ is the Boltzmann constant, and N is the number of magnetic ions per unit volume), we can deduce that the effective magnetic moment of the system increases with Mn doping. This can be attributed to the large magnetic moments of $Mn^{3+}$ and $Mn^{4+}$, and the substitution of nonmagnetic $Co^{3+}$ ions by $Mn^{3+}$ and $Mn^{4+}$ ions.

**Conclusions**

Single-phase layered $NaCo_{1-x}Mn_xO_2$ samples were prepared by the solid-state reaction. Upon Mn doping, the lattice parameter *a* decreases and the lattice parameter *c* increases rapidly, which lead to the expansion of the unit cell volume. $NaCo_{1-x}Mn_xO_2$ is very sensitive to Mn doping, as a result, substitution of 3% Mn for Co results in a metal-insulator transition. The temperature dependence of the magnetic moment for $NaCo_{1-x}Mn_xO_2$ samples obeys the Curie-Weiss law and the effective magnetic moment increases along with Mn doping. The results can be well explained by considering the valence states of Co and Mn, the variation of coupling between $CoO_2$ layers and the disorder induced by Mn doping.

**Acknowledgements**

This work was supported by NSFC (project No. 50272031), the Excellent Young Teacher Program of MOE, P.R.C, National 973 project (No. 2002CB613505) and Specialized Research Fund for the Doctoral Program of Higher Education (No. 2003 0003088).

**Figure captions**

Fig. 1 X-ray diffraction patterns of $NaCo_{1-x}Mn_xO_2$ with different doping concentrations.

Fig. 2 Variation of the lattice parameters *a*, c and unit cell volume with Mn content x.

Fig. 3 (a) EELS spectra of $NaCo_{0.9}Mn_{0.1}O_2$ (a) and $NaCo_{0.5}Mn_{0.5}O_2$ (b).

Fig. 3 (b) Typical spectra for the Mn $L_2$ and $L_3$ peaks obtained from x=0.1 (a) and 0.5 (b) samples, respectively.

Fig. 4 The variation of resistivity with temperature for $NaCo_{1-x}Mn_xO_2$. The inset is the dependence of room-temperature resistivity on Mn content x.

Fig. 5 Temperature dependence of magnetization for $NaCo_{1-x}Mn_xO_2$ samples.

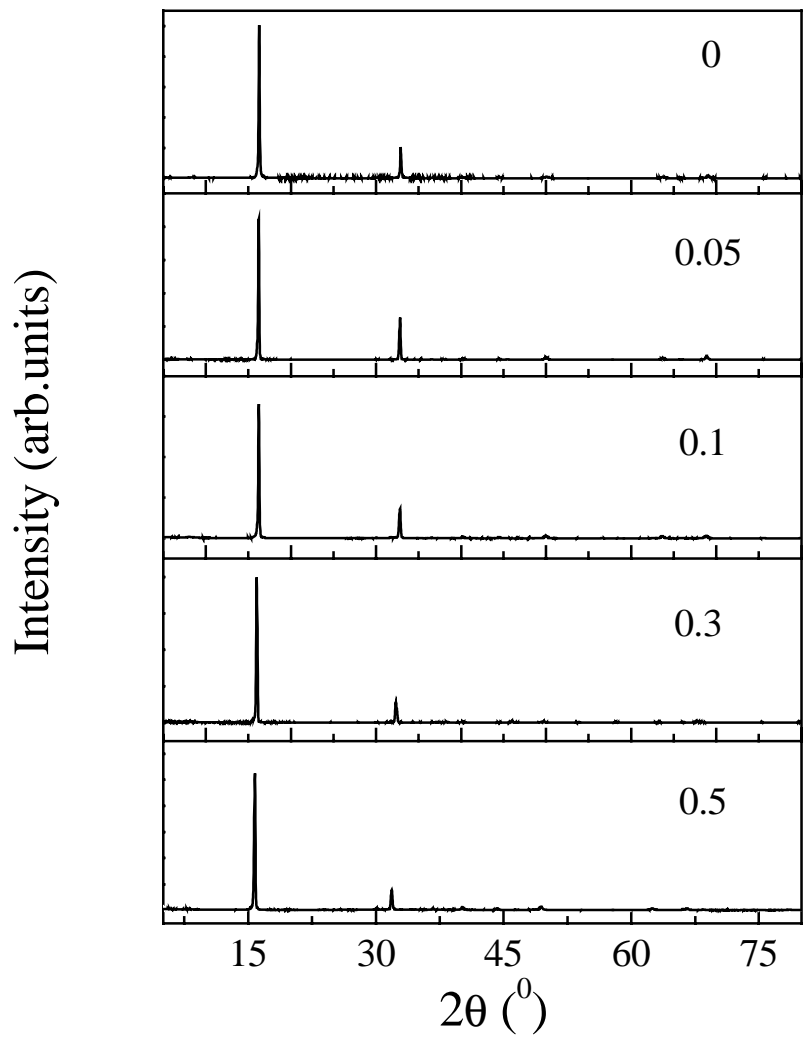

Fig.1

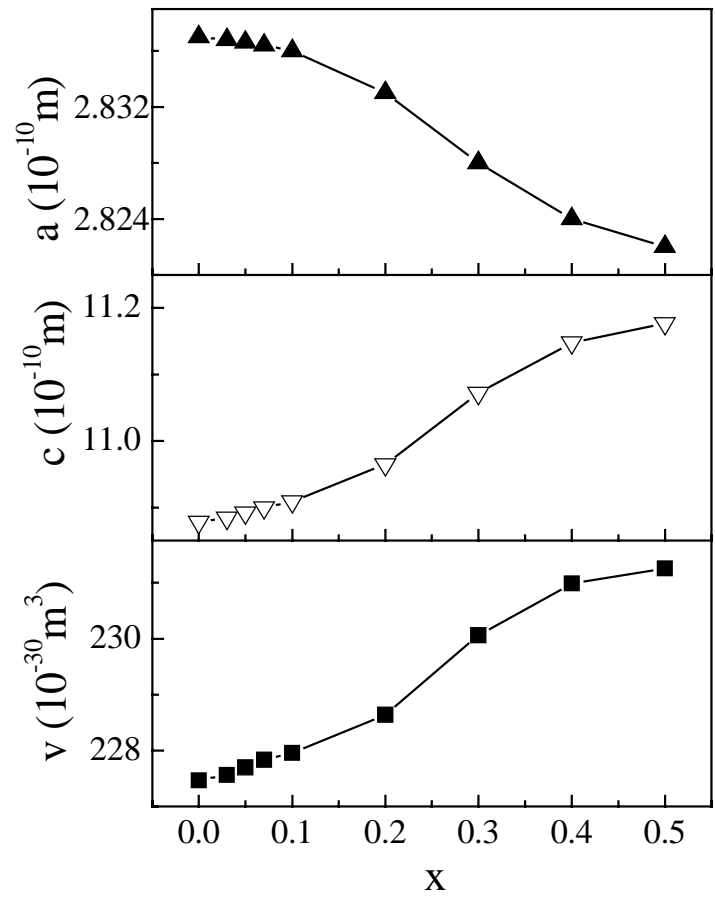

Fig.2

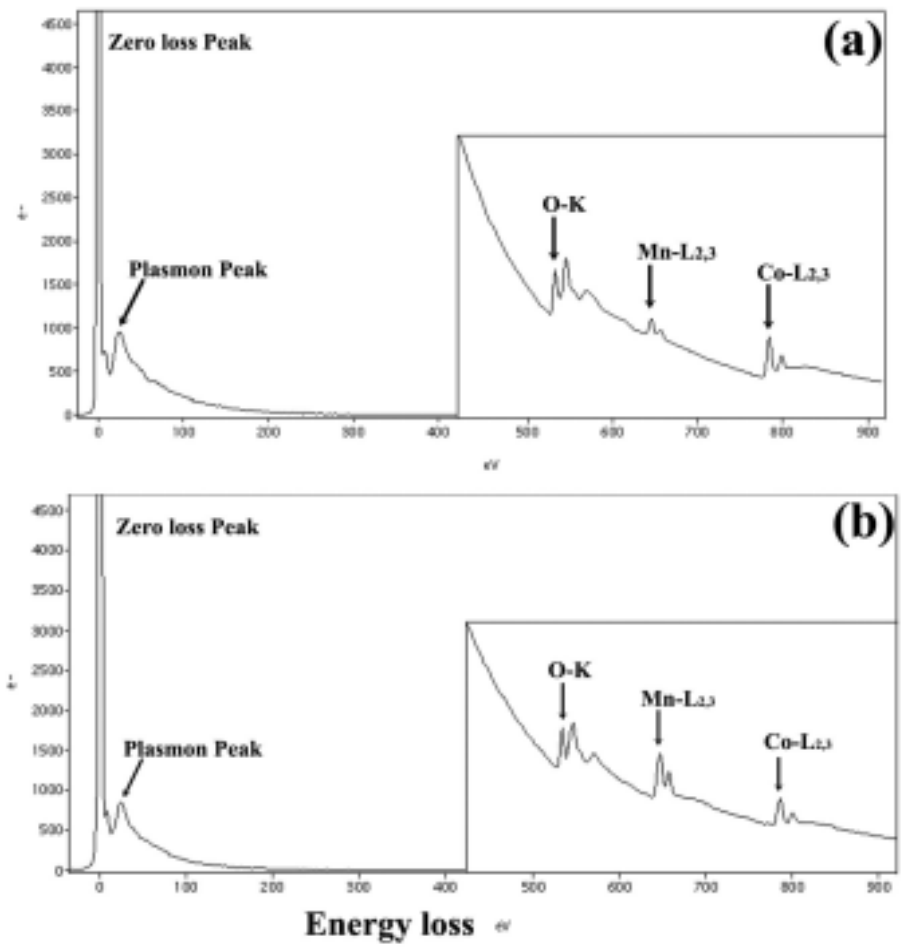

Fig. 3(a)

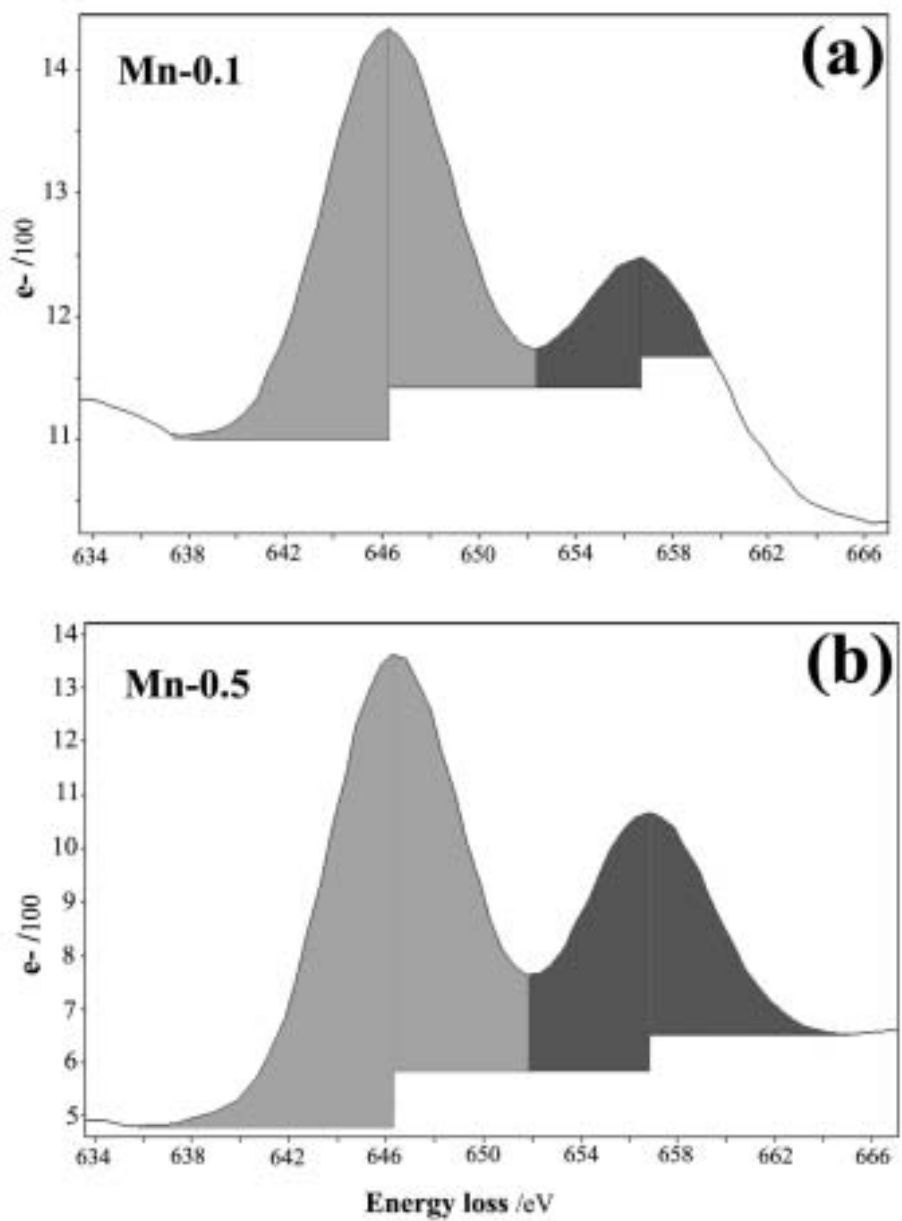

Fig. 3(b)

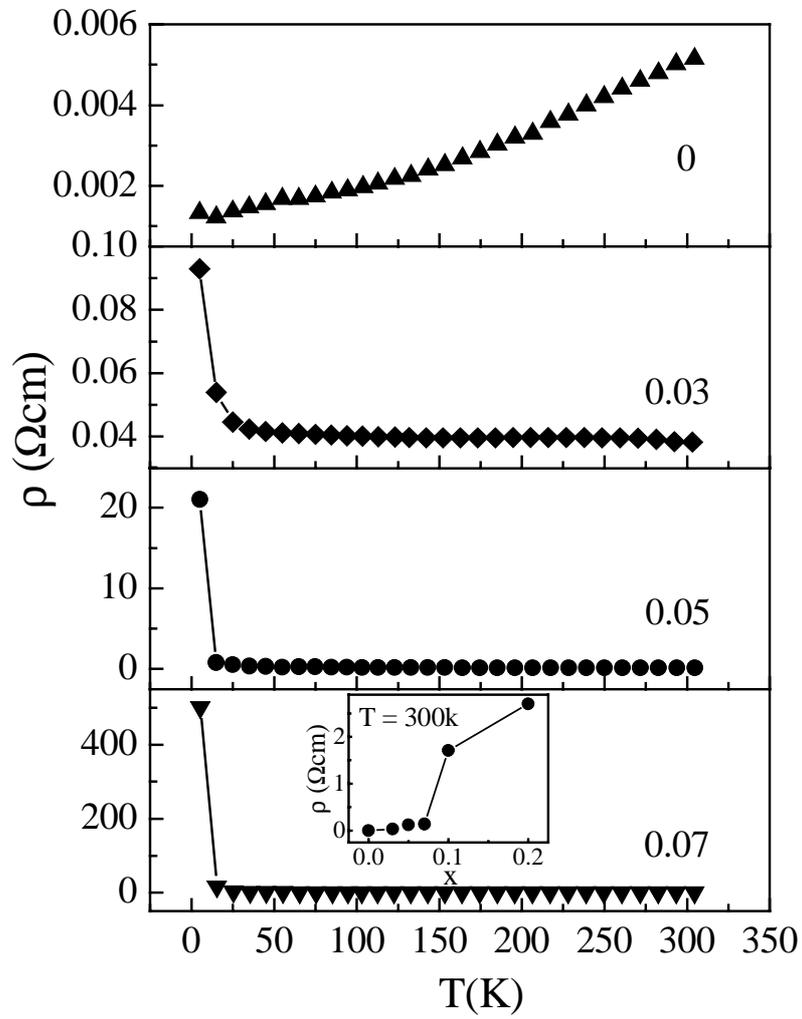

Fig.4

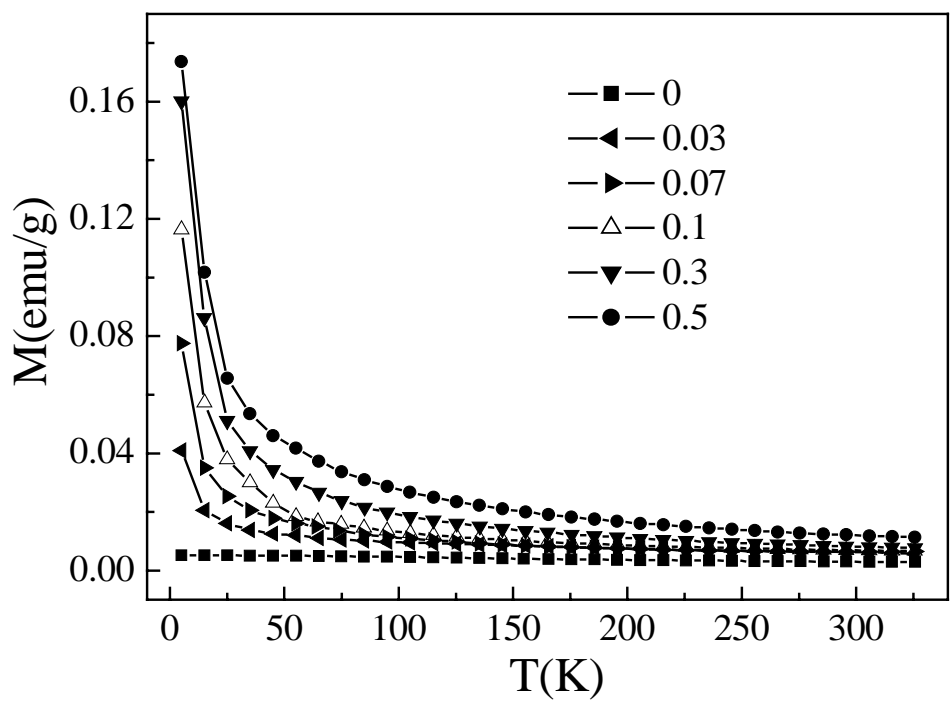

Fig5